\documentclass[preprint,superscriptaddress,aps]{revtex4}
\usepackage{amsfonts}
\usepackage{amssymb}
\usepackage{amsmath}
\usepackage{graphics}
\usepackage{graphicx}
\usepackage{color}
\usepackage[T1]{fontenc}
\usepackage{hyperref}
\newcommand{\bea}{\begin{eqnarray}}
\newcommand{\eea}{\end{eqnarray}}

\begin{document}
\title{Superfield Effective Potential for the 2-form field}

\author{C. A. S. Almeida}
\email{carlos@fisica.ufc.br}
\affiliation{Departamento de F\'{\i}sica, Universidade Federal do Cear\'{a} (UFC),\\
 Caixa Postal 6030, 60455-760, Fortaleza, CE, Brazil}

\author{F. S. Gama}
\email{fgama@fisica.ufpb.br}
\affiliation{Departamento de F\'{\i}sica, Universidade Federal da Para\'{\i}ba (UFPB),\\
 Caixa Postal 5008, 58051-970, Jo\~ao Pessoa, Para\'{\i}ba, Brazil}

\author{R. V. Maluf}
\email{r.v.maluf@fisica.ufc.br}
\affiliation{Departamento de F\'{\i}sica, Universidade Federal do Cear\'{a} (UFC),\\
 Caixa Postal 6030, 60455-760, Fortaleza, CE, Brazil}

\author{J. R. Nascimento}
\email{jroberto@fisica.ufpb.br}
\affiliation{Departamento de F\'{\i}sica, Universidade Federal da Para\'{\i}ba (UFPB),\\
 Caixa Postal 5008, 58051-970, Jo\~ao Pessoa, Para\'{\i}ba, Brazil}

\author{A. Yu. Petrov}
\email{petrov@fisica.ufpb.br}
\affiliation{Departamento de F\'{\i}sica, Universidade Federal da Para\'{\i}ba (UFPB),\\
 Caixa Postal 5008, 58051-970, Jo\~ao Pessoa, Para\'{\i}ba, Brazil}


\begin{abstract}
We develop a theory describing the superfield extension of the 2-form field coupled to usual chiral and real scalar superfield and find the one-loop K\"{a}hlerian effective potential in this theory.

\end{abstract}

\maketitle

\section{Introduction}

The supersymmetric field theory models are constructed on the base of specific supermultiplets represented by corresponding superfields \cite{BuKu,SGRS}. The most used supermultiplets are the chiral one widely used for description of the scalar matter, and the vector one naturally describing supersymmetric extensions of gauge theories. However, the set of possible supermultiplets is much larger. The most important their examples are presented in \cite{SGRS}.

One of important although less studied multiplets is the tensor one described by the spinor chiral superfield. Originally, it has been introduced in \cite{Siegel} where it was shown to describe a gauge theory. Further, it was demonstrated in \cite{GLA} that this superfield allows to construct the supersymmetric extension of the BF gauge theory in four-dimensional space-time, allowing thus for the superfield description of the models involving the antisymmetric tensor field which is essentially important within the string theory context \cite{SeWitt}, as well as within the quantum gravity context \cite{Frei}. While in \cite{GLA}, the free action for this theory was constructed, it is natural to make the next step, that is, to couple this theory to the matter, which is as usual represented by chiral scalar superfield, and to study the low-energy effective action in the resulting theory.

In our previous work \cite{prev}, the coupling of the spinor chiral gauge superfield to the chiral matter has been considered, and the leading one-loop contribution to the effective potential has been calculated. However, the action considered in \cite{prev} does not involve the terms responsible for the BF action. Therefore, we propose another theory which, from one side, is similar in some aspects to the model discussed in \cite{prev}, from another side, involves the BF terms, allowing thus to treat the BF theory in a manner analogous to \cite{prev}.

Within our studies, we consider the composite theory whose action involves, first, the usual superfield Maxwell term describing the dynamics of the real scalar gauge superfield, second, the action for the spinor chiral superfield involving the gauge invariant BF term, third, the coupling of these gauge fields to chiral matter. For this theory, we calculate the low-energy effective action described by the K\"{a}hlerian effective potential.

The structure of the paper reads as follows. In the section 2 we formulate the model involving two gauge fields and matter. In the section 3 we perform the one-loop calculations. In the summary, we discuss the results.

\section{The model}

We start with the Abelian gauge theory describing two gauge fields, the real scalar one $V$ and the chiral spinor one $\psi_{\alpha}$:
\bea
\label{kinetic}
S_k=\frac{1}{2}\int d^6z W^{\alpha}W_{\alpha}-\frac{1}{2}\int d^8zG^2 \ ,
\eea
where
\bea
\label{WG}
W_\alpha=i\bar D^2D_\alpha V \ , \ G=-\frac{1}{2}(D^\alpha\psi_\alpha+\bar D^{\dot{\alpha}}\bar\psi_{\dot{\alpha}}) \ .
\eea
The theory (\ref{kinetic}) is gauge invariant, with the corresponding gauge transformations look like:
\bea
\label{gaugetrans}
\delta V=i(\bar\Lambda-\Lambda) \ , \ \delta\psi_\alpha=i\bar D^2D_\alpha L \ , \ \delta\bar\psi_{\dot{\alpha}}=-iD^2\bar D_{\dot{\alpha}}L \ ,
\eea
where, as in \cite{SGRS}, the parameters $\Lambda$ and $\bar\Lambda$ are chiral and antichiral respectively, and $L=\bar L$ is a real one.

We can introduce mass terms for the theory (\ref{kinetic}). They are given by \cite{Siegel}
\bea
\label{massterm}
S_m&=&\frac{im}{2}\bigg[\int d^6z\psi^\alpha W_\alpha-\int d^6\bar z\bar\psi^{\dot{\alpha}}\bar W_{\dot{\alpha}}\bigg]+\frac{m^2_\psi}{4}\bigg[\int d^6z\psi^\alpha \psi_\alpha+\int d^6\bar z\bar\psi^{\dot{\alpha}}\bar\psi_{\dot{\alpha}}\bigg]\nonumber\\
&+&\frac{m_V^2}{2}\int d^8zV^2 \ .
\eea
Actually, the Eq. (\ref{massterm}), considered at $m_V=0$, describes the superfield BF model \cite{GLA}. In that paper, the dimensional reduction of that model has been carried out, and a mass generation mechanism for the Kalb-Ramond field was performed without loss of gauge and supersymmetry invariance.

Therefore, let us now consider the theory whose action is given by $S=S_k+S_m$. We note that the term $\int d^8z G^2$ in its action is necessary since if this term would absent, one could simply eliminate the $\psi_{\alpha}$ and $\bar{\psi}_{\dot{\alpha}}$ through their equations of motion thus reducing the theory to a simple supersymmetric QED.

Now, we can obtain the equations of motion for the model given by a sum of (\ref{kinetic})  and (\ref{massterm}) . For the superfields $V$, $\psi_\alpha$, and $\bar\psi_{\dot{\alpha}}$ respectively, they look like
\bea
\label{eqW}
\frac{\delta(S_k+S_m)}{\delta V}&=&iD_\alpha W^\alpha-mG+m_V^2V=0 \ ,\\
\label{eqG1}
2\frac{\delta(S_k+S_m)}{\delta \psi_\alpha}&=&\bar D^2D^\alpha G-imW^\alpha-m_\psi^2\psi^\alpha=0 \ ,\\
\label{eqG2}
2\frac{\delta(S_k+S_m)}{\delta\bar\psi_{\dot{\alpha}}}&=&D^2\bar D^{\dot{\alpha}} G+im\bar W^{\dot{\alpha}}-m_\psi^2\bar\psi^{\dot{\alpha}}=0 \ .
\eea
It follows from (\ref{eqW},\ref{eqG1},\ref{eqG2}) that, first, the superfield strengths $W_\alpha$ and $G$ satisfy the field equations \cite{prev}:
\bea
\label{KGSF}
(\Box-m^2)W^\alpha=0 \ , \ (\Box-m^2)G=0 \ ; \ \textrm{for} \ m_V=m_\psi=0 \ ,
\eea
second, the gauge superfields $V$, $\psi_\alpha$, and $\bar\psi_{\dot{\alpha}}$ satisfy the field equations \cite{BuKu}:
\bea
\label{KGGS}
(\Box-m_V^2)V=0 \ , \ (\Box-m_\psi^2)\psi^\alpha=0 \ , \ (\Box-m_\psi^2)\bar\psi^{\dot{\alpha}}=0 \ ; \ \textrm{for} \ m=0 \ .
\eea
We conclude that both the superfield strengths and the gauge superfields satisfy Klein-Gordon equations. Notice that (\ref{massterm}) is not invariant under the gauge transformations (\ref{gaugetrans}) unless $m_V=m_\psi=0$. As a consequence of this fact, the Eqs. (\ref{KGSF}) are invariant under the gauge transformations, but the Eqs. (\ref{KGGS}) are not.

In order to overcome the lack of gauge symmetry of the theory $S_k+S_m$, for $m_V\ne0$ and $m_\psi\ne0$, let us generalize (\ref{massterm}) by introducing the  St\"{u}ckelberg superfields $\Omega$, $\bar\Omega$, and $N$ in the following way:
\bea
\label{primemassterm}
S_m^\prime&=&\frac{im}{2}\bigg[\int d^6z\psi^\alpha W_\alpha-\int d^6\bar z\bar\psi^{\dot{\alpha}}\bar W_{\dot{\alpha}}\bigg]+\frac{m^2_\psi}{4}\bigg\{\int d^6z\Big[\psi^\alpha-\frac{i}{m_\psi}\bar D^2D^\alpha N\Big]\nonumber\\
&\times&\Big[\psi_\alpha-\frac{i}{m_\psi}\bar D^2D_\alpha N\Big]+\int d^6\bar z\Big[\bar\psi^{\dot\alpha}+\frac{i}{m_\psi}D^2\bar D^{\dot\alpha} N\Big]\Big[\bar\psi_{\dot\alpha}+\frac{i}{m_\psi} D^2\bar D_{\dot\alpha} N\Big]\bigg\}\nonumber\\
&+&\frac{m_V^2}{2}\int d^8z\Big[V+\frac{i}{m_V}(\Omega-\bar\Omega)\Big]^2 \ ,
\eea
where these new superfields transform as
\bea
\label{Stutrans}
\delta \Omega=m_V\Lambda \ , \ \delta \bar\Omega=m_V\bar\Lambda \ , \ \delta N=m_\psi L \ .
\eea
By construction, the action (\ref{primemassterm}) (and $S_k+S_m^\prime$) is invariant under the gauge transformations (\ref{gaugetrans}) and (\ref{Stutrans}).

Note that there are mixed St\"{u}ckelberg and gauge superfield terms in (\ref{primemassterm}). This makes the one-loop calculations more cumbersome. However, if one fixes the gauge through adding the gauge-fixing term of the form:
\bea
\label{gaugefix}
S_{GF}&=&-\frac{1}{\alpha}\int d^8z\Big(\bar D^2V-i\alpha m_V\frac{\bar D^2}{\Box}\bar\Omega\Big)\Big(D^2V+i\alpha m_V\frac{D^2}{\Box}\Omega\Big)-\frac{1}{8\beta}\int d^8z(D^\alpha\psi_\alpha\nonumber\\
&-&\bar D^{\dot{\alpha}}\bar\psi_{\dot{\alpha}}+2i\beta m_\psi N)^2 \ ,
\eea
where $\alpha$ and $\beta$ are the gauge-fixing parameters, the mixed terms are eliminated. Of course, since the gauge symmetry in this theory is Abelian, the ghosts completely decouple.

Up to now, we considered only the free theory. Now, let us introduce its coupling to the matter represented as usual by chiral and antichiral scalar fields \cite{BuKu}.
It is known that under the usual gauge transformation, the chiral and antichiral matter superfields transform as \cite{SGRS}
\bea
\label{mattertrans}
\Phi^{\prime}=e^{2ig\Lambda}\Phi \ , \  \bar\Phi^{\prime}=\bar\Phi e^{-2ig\bar\Lambda} \ .
\eea
Then, we introduce the following gauge invariant action involving coupling of matter and gauge fields
\cite{Chris} studied also  in \cite{Fe} within the cosmic strings context:
\bea
\label{matteract}
S_M=\int d^8z\bar\Phi e^{2gV}\Phi e^{4hG} \ .
\eea
The coupling constants $g$ and $h$ have mass dimensions $0$ and $-1$, respectively (the last fact implies the non-renormalizability of the theory; however, renormalizable and gauge invariant couplings of superfields $\psi_{\alpha}$ and $\bar{\psi}_{\dot{\alpha}}$ simply do not exist).

Finally, the complete supersymmetric massive gauge theory we study here is described by the sum of (\ref{kinetic}), (\ref{primemassterm}) (\ref{gaugefix}), and (\ref{matteract}), that is:
\bea
\label{totaltheory}
S&=&-\frac{1}{2}\int d^8z V(-D^\alpha \bar D^2D_\alpha+\frac{1}{\alpha}\{D^2,\bar D^2\})V-\frac{1}{8}\int d^8z \Big\{\big(1+\frac{1}{\beta}\big)[\psi_\alpha D^\alpha D^\beta\psi_\beta\nonumber\\
&&+\bar\psi_{\dot{\alpha}}\bar D^{\dot{\alpha}}\bar D^{\dot{\beta}}\bar\psi_{\dot{\beta}}]+2\big(1-\frac{1}{\beta}\big)\psi_\alpha D^\alpha\bar D^{\dot{\beta}}\bar\psi_{\dot{\beta}}\Big\}+\frac{m}{2}\int d^8zV(D^\alpha\psi_\alpha+\bar D^{\dot{\alpha}}\bar\psi_{\dot{\alpha}})\nonumber\\
&&+\frac{m_V^2}{2}\int d^8zV^2+\frac{1}{2}\int d^8z\bigg[(D^\alpha\psi_\alpha)\frac{m^2_\psi}{2\Box}(D^\beta\psi_\beta)+(\bar{D}^{\dot{\alpha}}\bar\psi_{\dot{\alpha}})\frac{m^2_\psi}{2\Box}
(\bar{D}^{\dot{\beta}}\bar\psi_{\dot{\beta}})\bigg]\nonumber\\
&&+\int d^8z\bar\Phi e^{2gV}\Phi e^{-2h(D^\alpha\psi_\alpha+\bar D^{\dot{\alpha}}\bar\psi_{\dot{\alpha}})} \ +(\ldots),
\eea
where the dependence on the gauge superfields is given explicitly. Here the dots are for the contributions involving the St\"{u}ckelberg superfields which completely decouple giving only a trivial contribution to the
effective action. Finally, notice that there is a nonlocality which was introduced in order
to rewrite the mass term as an integral over whole superspace.

Now, let us calculate the effective action for our theory. It is known \cite{BuKu} that in the matter sector, the low-energy effective action in theories involving chiral and antichiral matter fields is characterized by the K\"ahlerian effective potential (KEP) depending only on the background matter fields but not on their derivatives. Within this paper we concentrate namely in calculating the KEP $K(\Phi,\bar\Phi)$ in our theory.

The standard method of calculating the effective action is based on the methodology of the loop expansion \cite{ourcourse,BO}. To do this, we make a shift $\Phi\rightarrow\Phi+\phi$ in the superfield $\Phi$ (together with the analogous shift for $\bar\Phi$), where now $\Phi$ is a background (super)field and $\phi$ is a quantum one.
Since our aim in this paper will consist in consideration of the KEP,
we assume that the gauge superfields $V$, $\psi_\alpha$, and $\bar\psi_{\dot{\alpha}}$ are purely quantum ones. In the one-loop approximation, one should keep only the quadratic terms in the quantum superfields. Therefore, (\ref{totaltheory}) implies the following quadratic action of quantum superfields:
\bea
\label{backaction1}
&&S_2[\bar\Phi,\Phi;\bar\phi,\phi,\psi_\alpha,\bar\psi_{\dot{\alpha}},V]=S_q+S_{int} \ ,\\
&&S_q=\frac{1}{2}\int d^8z\big[ -V\Box(\Pi_{1/2}+\frac{1}{\alpha}\Pi_0)V-\frac{1}{2}\psi_\alpha D^\alpha\bar D^{\dot{\beta}}\bar\psi_{\dot{\beta}}+2\bar\phi\phi\big] \ ,\\
\label{intver}
&&S_{int}=\frac{1}{2}\int d^8z\big\{(m-8gh\bar\Phi\Phi)V(D^\alpha\psi_\alpha+\bar D^{\dot{\alpha}}\bar\psi_{\dot{\alpha}})+2(2g)\bar\Phi V\phi+2(2g)\Phi\bar\phi V\nonumber\\
&& \ \ \ \ +\big(m^2_V+(2g)^2\bar\Phi\Phi\big)V^2-4h\bar\Phi(D^\alpha\psi_\alpha+\bar D^{\dot{\alpha}}\bar\psi_{\dot{\alpha}})\phi-4h\Phi\bar\phi(D^\alpha\psi_\alpha+\bar D^{\dot{\alpha}}\bar\psi_{\dot{\alpha}})\nonumber\\
&& \ \ \ \ +(D^\alpha\psi_\alpha)\Big[-\frac{1}{4}\big(1+\frac{1}{\beta}\big)+\frac{m_\psi^2}{2\Box}+(2h)^2\bar\Phi\Phi\Big]D^\beta\psi_\beta+(\bar D^{\dot{\alpha}}\bar\psi_{\dot{\alpha}})\Big[-\frac{1}{4}\big(1+\frac{1}{\beta}\big)\nonumber\\
&& \ \ \ \ +\frac{m_\psi^2}{2\Box}+(2h)^2\bar\Phi\Phi\Big]\bar D^{\dot{\beta}}\bar\psi_{\dot{\beta}}+2\Big[\frac{1}{4\beta}+(2h)^2\bar\Phi\Phi\Big](D^\alpha\psi_\alpha)\bar D^{\dot{\alpha}}\bar\psi_{\dot{\alpha}}\big\} \ ,
\eea
where the terms involving derivatives of the background superfields were omitted being irrelevant for us. Here, we use the projection operators $\Pi_{1/2}\equiv-\Box^{-1}D^\alpha\bar D^2D_\alpha$ and $\Pi_{0}\equiv\Box^{-1}\{D^2,\bar D^2\}$.

The one-loop approximation does not depend on the manner of splitting the Lagrangian into free and interacting parts since, at this order, one should deal only with the quadratic action of quantum superfields \cite{Coleman}. Usually, the propagators are defined from the background-independent terms, and the vertices are defined from the ones involving couplings of quantum superfields with the background ones. However, as a matter of convenience, here we will extract the propagators from $S_q$ and treat the remaining terms as interaction vertices. Therefore, we obtain from $S_q$ the propagators
\bea
\label{propagators}
\langle V(1)V(2)\rangle=-\frac{1}{p^2}(\Pi_{1/2}+\alpha\Pi_0)_1\delta_{12} \ , \
\langle\psi_\alpha(1)\bar\psi_{\dot{\alpha}}(2)\rangle=\frac{4p_{\alpha\dot{\alpha}}}{p^4}\delta_{12} \ , \
\langle\phi(1)\bar\phi(2)\rangle=\frac{1}{p^2}\delta_{12} \ .
\eea
It is convenient to transfer the covariant derivatives from the vertices to the propagators of the 2-form superfield. This will allow us to define new scalar propagators written in terms of projection operators. To do it, let us employ some tricks used in \cite{prev}: one can observe from (\ref{intver}) that there is a factor $D^\alpha\bar D^2$ in a vertex associated to one end of the propagator $\langle\psi_\alpha(1)\bar\psi_{\dot{\alpha}}(2)\rangle$, and there is a factor $\bar D^{\dot{\alpha}}D^2$ in the other vertex at other end of the same propagator. Thus, we absorb these covariant derivatives into redefinition of the propagator $\langle\psi_\alpha(1)\bar\psi_{\dot{\alpha}}(2)\rangle$ (instead of associating them to vertices) and define a new scalar field $\psi=D^{\alpha}\psi_{\alpha}$ whose propagator is:
\bea
\langle\psi(1)\bar\psi(2)\rangle&\equiv& D^\alpha_1\bar D^2_1\bar D^{\dot{\alpha}}_2 D^2_2\langle\psi_\alpha(1)\bar\psi_{\dot{\alpha}}(2)\rangle=4(\Pi_{1/2})_1\delta_{12} \ ,
\eea
where we took into account that $\bar D^{\dot{\alpha}}_2 D^2_2\delta_{12}=-D^2_1\bar D^{\dot{\alpha}}_1\delta_{12}$. We proceed in the same way with vertices and propagators involving $\phi$ and $\bar{\phi}$.

Finally, redefining the propagators and vertices in this manner, we get
\bea
\label{propagator1}
\langle V(1)V(2)\rangle&=&-\frac{1}{p^2}(\Pi_{1/2}+\alpha\Pi_0)_1\delta_{12} \ , \\
\label{propagator2}
\langle\psi(1)\bar\psi(2)\rangle&=&\langle\bar\psi(1)\psi(2)\rangle=4(\Pi_{1/2})_1\delta_{12} \ , \\
\label{propagator3}
\langle\phi(1)\bar\phi(2)\rangle&=&-(\Pi_-)_1\delta_{12} \ , \ \langle\bar\phi(1)\phi(2)\rangle=-(\Pi_+)_1\delta_{12} \ ,
\eea
where $\Pi_-\equiv\Box^{-1}\bar D^2D^2$ and $\Pi_+\equiv\Box^{-1}D^2\bar D^2$ are projection operators. The new interaction part of the action looks like
\bea
\label{newintver}
\tilde S_{int}&=&\frac{1}{2}\int d^8z\big\{2MV(\psi+\bar\psi)+2(2g)\bar\Phi V\phi+2(2g)\Phi\bar\phi V+\big(m^2_V+(2g)^2\bar\Phi\Phi\big)V^2\\
&&+\psi\Big[-\frac{1}{4}\big(1+\frac{1}{\beta}\big)+M_\psi\Big]\psi+\bar\psi\Big[-\frac{1}{4}\big(1+\frac{1}{\beta}\big) +M_\psi\Big]\bar\psi+2\Big[\frac{1}{4\beta}+(2h)^2\bar\Phi\Phi\Big]\psi\bar\psi\big\}\nonumber \ ,
\eea
where $M\equiv\frac{1}{2}(m-8gh\bar\Phi\Phi)$ and $M_\psi\equiv\frac{m_\psi^2}{2\Box}+(2h)^2\bar\Phi\Phi$ (notice that $M_{\psi}$, although characterizes the mass term, has zero mass dimension). We note that we redefined the theory in terms of scalar superfields only, which allows to simplify the calculations drastically.

In the next section, namely the new propagators (\ref{propagator1}-\ref{propagator3}) and the new vertices (\ref{newintver}), written only in terms of scalar superfields, will be used.

\section{One-loop calculations}

So, let us proceed with the calculating the KEP. The usual methods of its calculation are performed by means of perturbative series in powers of $\hbar$, the so-called loop expansion \cite{BuKu,ourcourse}, namely
\bea
K(\Phi,\bar\Phi)=K^{(0)}(\Phi,\bar\Phi)+\hbar K^{(1)}(\Phi,\bar\Phi)+\hbar^2 K^{(2)}(\Phi,\bar\Phi)+\cdots \ .
\eea
The tree approximation can be read from the classical action (\ref{matteract}) by replacing $g$ and $h$ by zero, yielding
\bea
K^{(0)}(\Phi,\bar\Phi)=\Phi\bar{\Phi} \ .
\eea
In order to calculate the one-loop contribution $K^{(1)}(\Phi,\bar\Phi)$, we will use the methodology of summation over supergraphs originally elaborated in \cite{SYM} and applied in many other examples including \cite{prev}.

We proceed in three steps. First, we draw all one-loop supergraphs allowed by (\ref{newintver}). Second, we discard supergraphs involving covariant derivatives of $\Phi$ and $\bar\Phi$, and calculate the contributions of each supergraph, with the external momenta equal to zero, to the effective action. Finally, we sum all contributions and calculate the integral over the momenta. The result will be just the KEP.

Due to the known properties of the supersymmetric projectors, that is, $\Pi_{1/2}\Pi_-=\Pi_-\Pi_{1/2}=\Pi_{1/2}\Pi_+=\Pi_+\Pi_{1/2}=0$, and to the fact that there is no any spinor covariant derivative in the vertices (\ref{newintver}), it follows from (\ref{propagator2}, \ref{propagator3}) that the mixed contributions containing both $\langle\psi(1)\bar\psi(2)\rangle$ and $\langle\phi(1)\bar\phi(2)\rangle$ propagators cannot arise. Therefore, the set of the one-loop supergraphs contributing to the effective action in the theory under consideration are of four types.

In our graphical notation, solid lines denote $\langle\phi(1)\bar\phi(2)\rangle$-propagators, the dashed ones denote $\langle\psi(1)\bar\psi(2)\rangle$-propagators, the wavy ones denote $\langle V(1)V(2)\rangle$-propagators, and the double ones denote $\Phi$ or $\bar{\Phi}$ background superfields.

Let us start the calculations of the one-loop supergraphs involving only the gauge superfield propagators $\langle V(1)V(2)\rangle$ in the internal lines connecting the vertices $\big(m^2_V+(2g)^2\bar\Phi\Phi\big)V^2$. Such supergraphs are the simplest and exhibit structures given at Fig. 1. Of course, if we had taken the particular case $\alpha=1$, such one-loop corrections would be zero, because the corrections would not contain any $D^2\bar D^2$ acting on the Grassmann delta-function.

\begin{figure}[!h]
\begin{center}
\includegraphics[angle=0,scale=0.40]{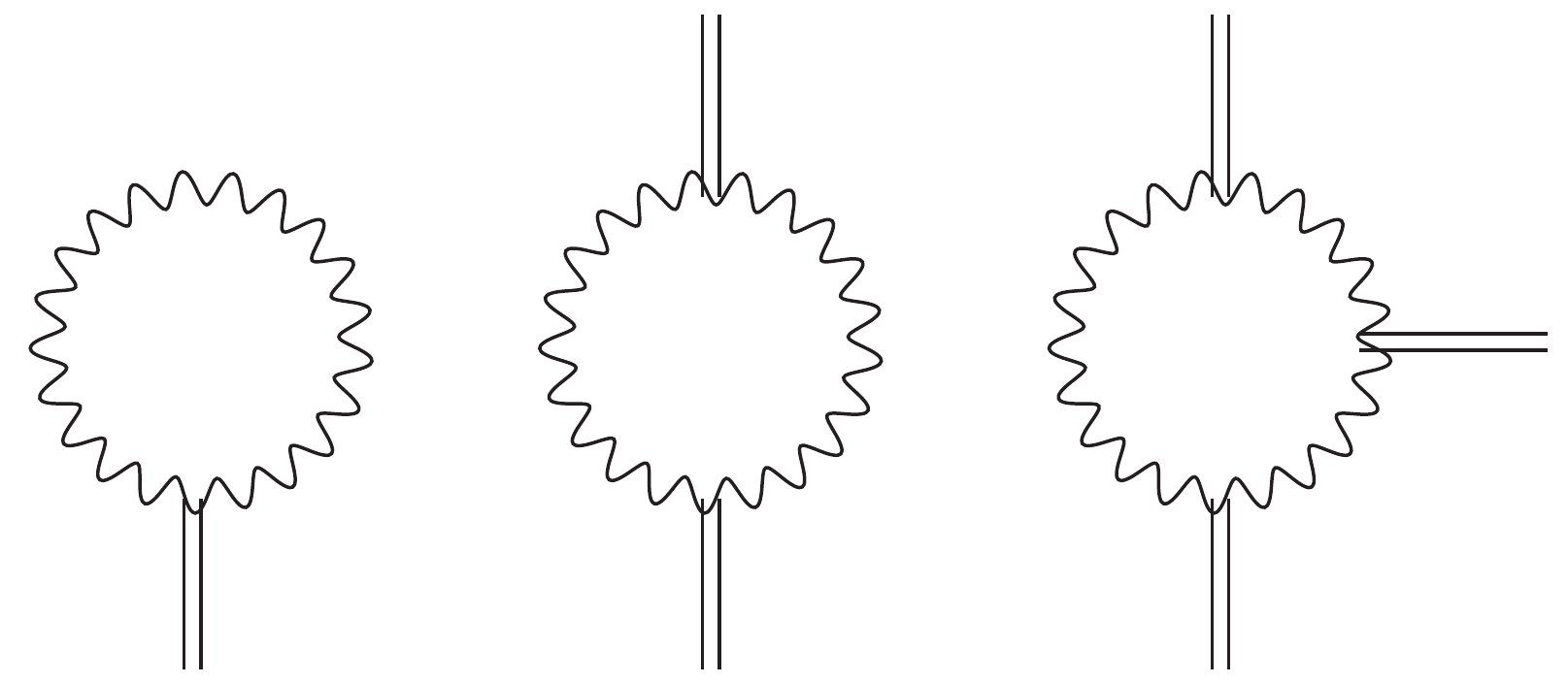}
\end{center}
\caption{One-loop supergraphs composed by propagators $\langle V(1)V(2)\rangle$.}
\end{figure}

We can compute all the contributions by noting that each supergraph above is formed by $n$ links, like those shown in Fig. 2.

\begin{figure}[!h]
\begin{center}
\includegraphics[angle=0,scale=0.60]{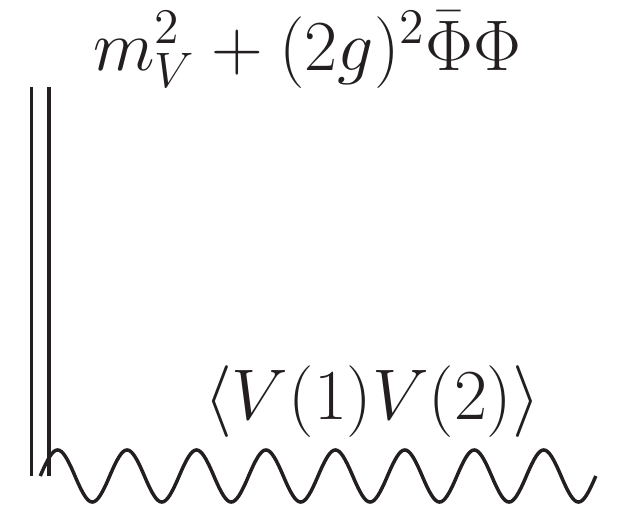}
\end{center}
\caption{A typical vertex in one-loop supergraphs involving $(m^2_V+(2g)^2\bar\Phi\Phi)V^2$.}
\end{figure}

Hence, the contribution of this link is simply given by
\bea
Q_{12}=\big(m^2_V+(2g)^2\bar\Phi\Phi\big)_1\Big[-\frac{1}{p^2}(\Pi_{1/2}+\alpha\Pi_0)\Big]_1 \delta_{12} \ .
\eea
Therefore, the contribution of a loop formed by $n$ such links is given by
\bea
(I_a)_n&=&\int d^4x\frac{1}{2n}\int d^4\theta_1d^4\theta_3\ldots d^4\theta_{2n-1}\int \frac{d^4p}{(2\pi)^4}Q_{13}Q_{35}\ldots Q_{2n-3,2n-1} Q_{2n-1,1} \nonumber\\
&=&\int d^8z\int \frac{d^4p}{(2\pi)^4}\frac{1}{2n}\bigg(-\frac{m^2_V+(2g)^2\bar\Phi\Phi}{p^2}\bigg)^{n}(\Pi_{1/2}+\alpha^n\Pi_0)\delta_{\theta\theta^{\prime}}|_{\theta=\theta^{\prime}}\ ,
\eea
where we integrated by parts the expression $(I_a)_n$ and used the usual properties of the projection operators.

The contribution for the effective action is given by the sum of all supergraphs $(I_a)_n$
\bea
\label{part1}
\Gamma^{(1)}_a&=&\sum_{n=1}^{\infty}(I_a)_n=\int d^8z\int \frac{d^4p}{(2\pi)^4}\frac{1}{p^2}\bigg\{-\ln\Big[1+\frac{m^2_V+(2g)^2\bar\Phi\Phi}{p^2}\Big]\nonumber\\
&+&\ln\bigg[1+\alpha\frac{m^2_V+(2g)^2\bar\Phi\Phi}{p^2}\bigg]\bigg\} \ .
\eea
Notice that this contribution vanishes at $\alpha=1$ (Feynman gauge), as it should be.

Let us proceed the calculation of the second type of one-loop supergraphs, which involve the $\langle\phi(1)\bar\phi(2)\rangle$ and $\langle V(1)V(2)\rangle$ propagators in the internal lines connecting the vertices $(2g)\Phi\bar\phi V$ and $(2g)\bar\Phi V\phi$. Such supergraphs exhibit the structure shown in Fig. 3. Certainly, if we had taken the particular case $\alpha=0$, such one-loop corrections would not contribute to the effective action, because $\langle V(1)V(2)\rangle\sim\Pi_{1/2}$ and $\Pi_{1/2}\Pi_-=\Pi_{1/2}\Pi_+=0$.
\begin{figure}[!h]
\begin{center}
\includegraphics[angle=0,scale=0.45]{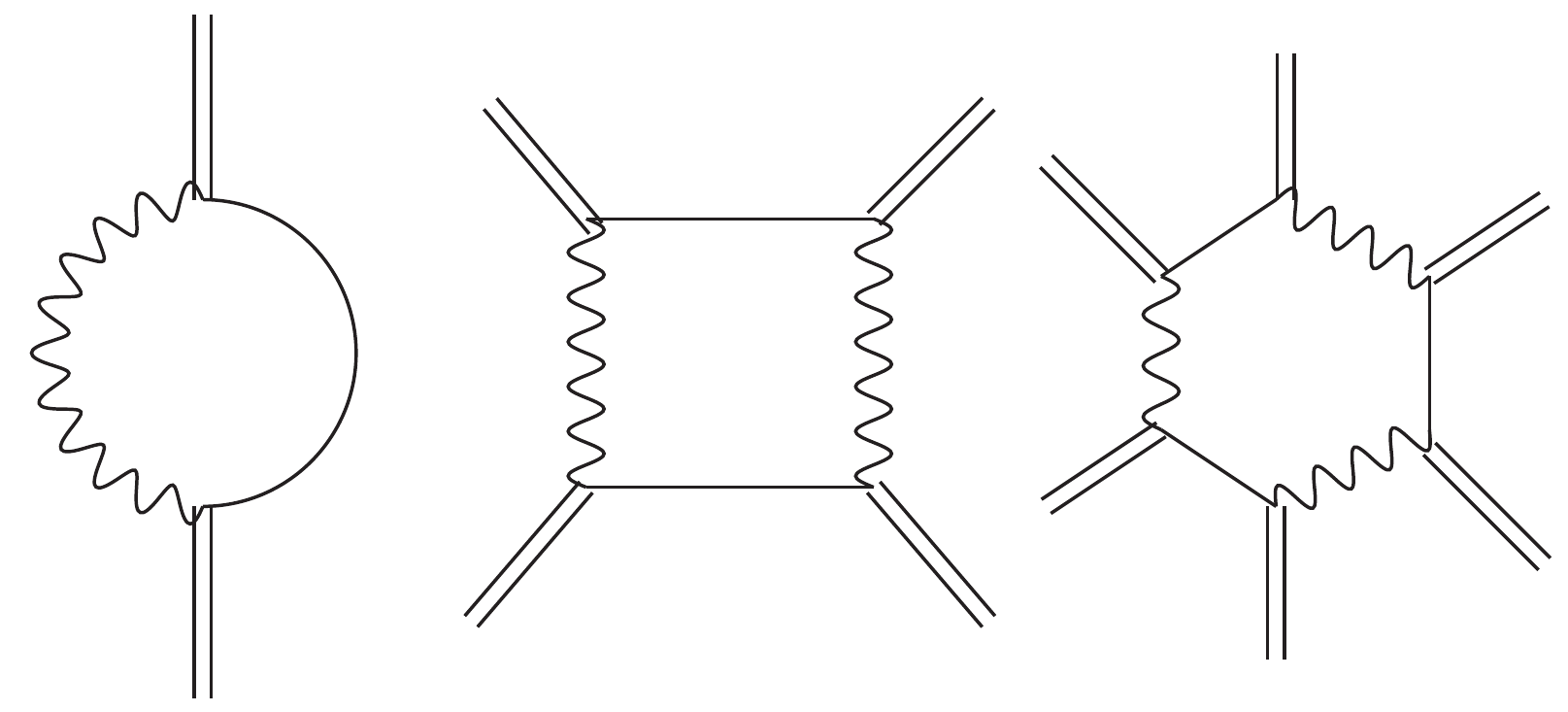}
\end{center}
\caption{One-loop supergraphs composed by propagators $\langle\phi(1)\bar\phi(2)\rangle$ and $\langle V(1)V(2)\rangle$.}
\end{figure}

\begin{figure}[!h]
\begin{center}
\includegraphics[angle=0,scale=0.60]{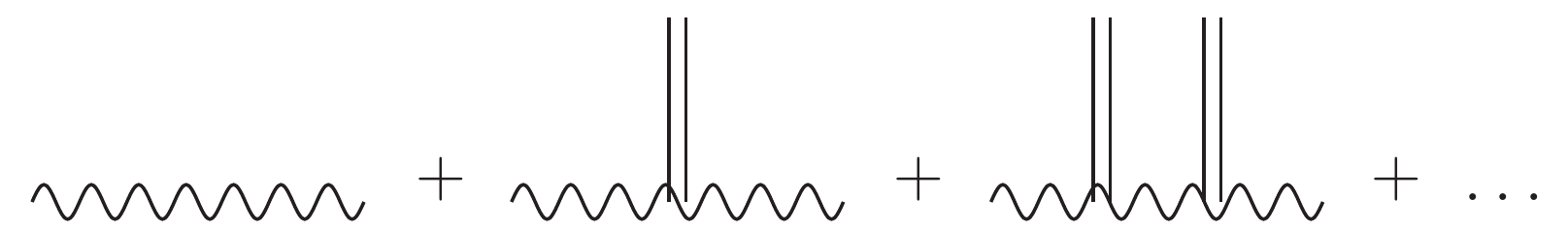}
\end{center}
\caption{Dressed propagator.}
\end{figure}

To sum over arbitrary numbers of insertions of vertices $\big(m^2_V+(2g)^2\bar\Phi\Phi\big)V^2$ into the gauge propagators, it is convenient to define a "dressed" propagator where the summation over all vertices $\big(m^2_V+(2g)^2\bar\Phi\Phi\big)V^2$ is performed (see Fig. 4), which, as a result, is equal to
\bea
\langle V(1)V(2)\rangle_D&=&\langle V(1)V(2)\rangle+\int d^4\theta_3\langle V(1)V(3)\rangle[m^2_V+(2g)^2\bar\Phi\Phi]_3\langle V(3)V(2)\rangle\nonumber\\
&+&\int d^4\theta_3d^4\theta_4\langle V(1)V(3)\rangle[m^2_V+(2g)^2\bar\Phi\Phi]_3\langle V(3)V(4)\rangle[m^2_V+(2g)^2\bar\Phi\Phi]_4\nonumber\\
&\times&\langle V(4)V(2)\rangle+\ldots \ .
\eea
Finally, we arrive at
\bea
\label{dressedpropagator}
\langle V(1)V(2)\rangle_D=-\bigg[\frac{\Pi_{1/2}}{p^2+m^2_V+(2g)^2\bar\Phi\Phi}+\frac{\alpha\Pi_0}{p^2+\alpha\big(m^2_V+(2g)^2\bar\Phi\Phi\big)}\bigg]_1\delta_{12} \ .
\eea
Then, we notice that each supergraph above (see Fig. 3) is formed by $n$ links depicted in Fig. 5, each of which yields the contribution
\bea
R_{13}&=&\int d^4\theta_2[(2g)\Phi]_1\big\{-
\bigg[\frac{\Pi_{1/2}}{p^2+m^2_V+(2g)^2\bar\Phi\Phi}+\frac{\alpha\Pi_0}{p^2+\alpha\big(m^2_V+(2g)^2\bar\Phi\Phi\big)}\bigg]_1\delta_{12}\big\}\nonumber\\
&\times&[(2g)\bar\Phi]_2\Big[-(\Pi_-)_2 \delta_{23}\Big]\nonumber\\
&=&\bigg[\frac{\alpha(2g)^2\bar\Phi\Phi}{p^2+\alpha\big(m^2_V+(2g)^2\bar\Phi\Phi\big)}\Pi_-\bigg]_1\delta_{13} \ .
\eea

\begin{figure}[!h]
\begin{center}
\includegraphics[angle=0,scale=0.60]{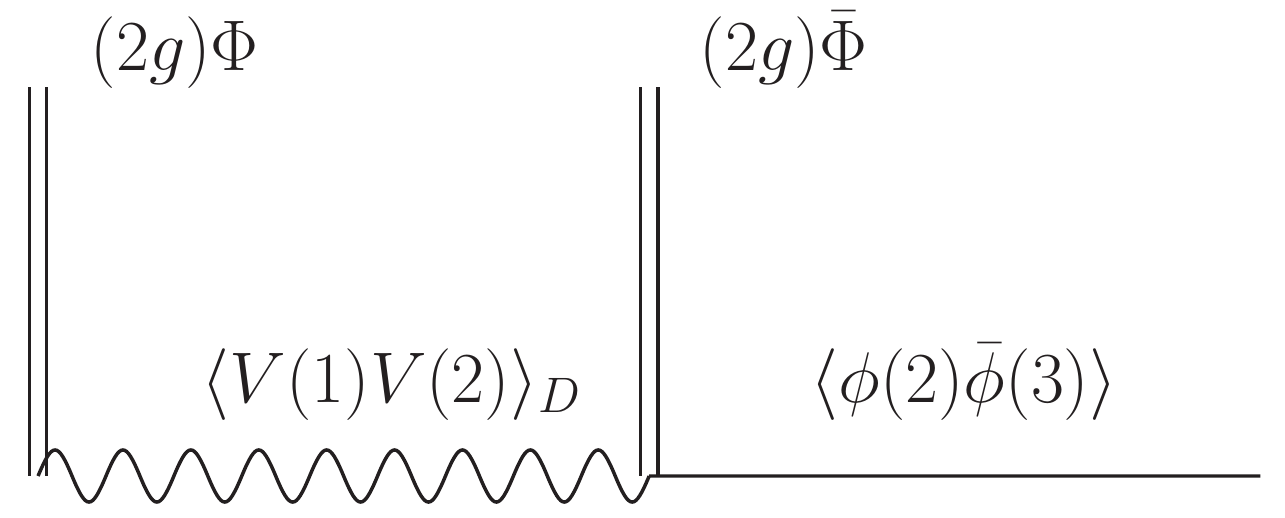}
\end{center}
\caption{A typical link in one-loop supergraphs in mixed sector.}
\end{figure}

Therefore the contribution of a supergraph formed by $n$ such links is given by
\bea
(I_b)_n&=&\int d^4x\frac{1}{n}\int d^4\theta_1d^4\theta_3\ldots d^4\theta_{2n-1}\int \frac{d^4p}{(2\pi)^4}R_{13}R_{35}\ldots R_{2n-3,2n-1} R_{2n-1,1} \nonumber\\
&=&\int d^8z\frac{1}{n}\int \frac{d^4p}{(2\pi)^4}\bigg(\frac{\alpha(2g)^2\bar\Phi\Phi}{p^2+\alpha\big(m^2_V+(2g)^2\bar\Phi\Phi\big)}\bigg)^n\Pi_-\delta_{\theta\theta^{\prime}}|_{\theta=\theta^{\prime}}\ .
\eea
By using $\Pi_-\delta_{\theta\theta^{\prime}}|_{\theta=\theta^{\prime}}=-1/p^2$, we get the effective action
\bea
\label{part2}
\Gamma^{(1)}_b=\sum_{n=1}^{\infty}(I_b)_n=\int d^8z\frac{1}{p^2}\ln\bigg[\frac{p^2+\alpha m^2_V}{p^2+\alpha\big(m^2_V+(2g)^2\bar\Phi\Phi\big)}\bigg] \ .
\eea
We notice that at $\alpha=0$ (Landau gauge), this expression vanishes.

By summing (\ref{part1}) and (\ref{part2}), we get
\bea
\label{part1+2}
\Gamma^{(1)}_a+\Gamma^{(1)}_b=\int d^8z\int \frac{d^4p}{(2\pi)^4}\frac{1}{p^2}\bigg\{-\ln\Big[1+\frac{m^2_V+(2g)^2\bar\Phi\Phi}{p^2}\Big]+\ln\bigg[1+\alpha\frac{m^2_V}{p^2}\bigg]\bigg\} \ .
\eea
Notice that (\ref{part1+2}) is explicitly gauge independent for the massless case $m_V=0$. However, even in the massive case, the $\alpha$-dependence is trivial since the last logarithm does not depend on the background superfields and hence can be disregarded.


\begin{figure}[!h]
\begin{center}
\includegraphics[angle=0,scale=0.40]{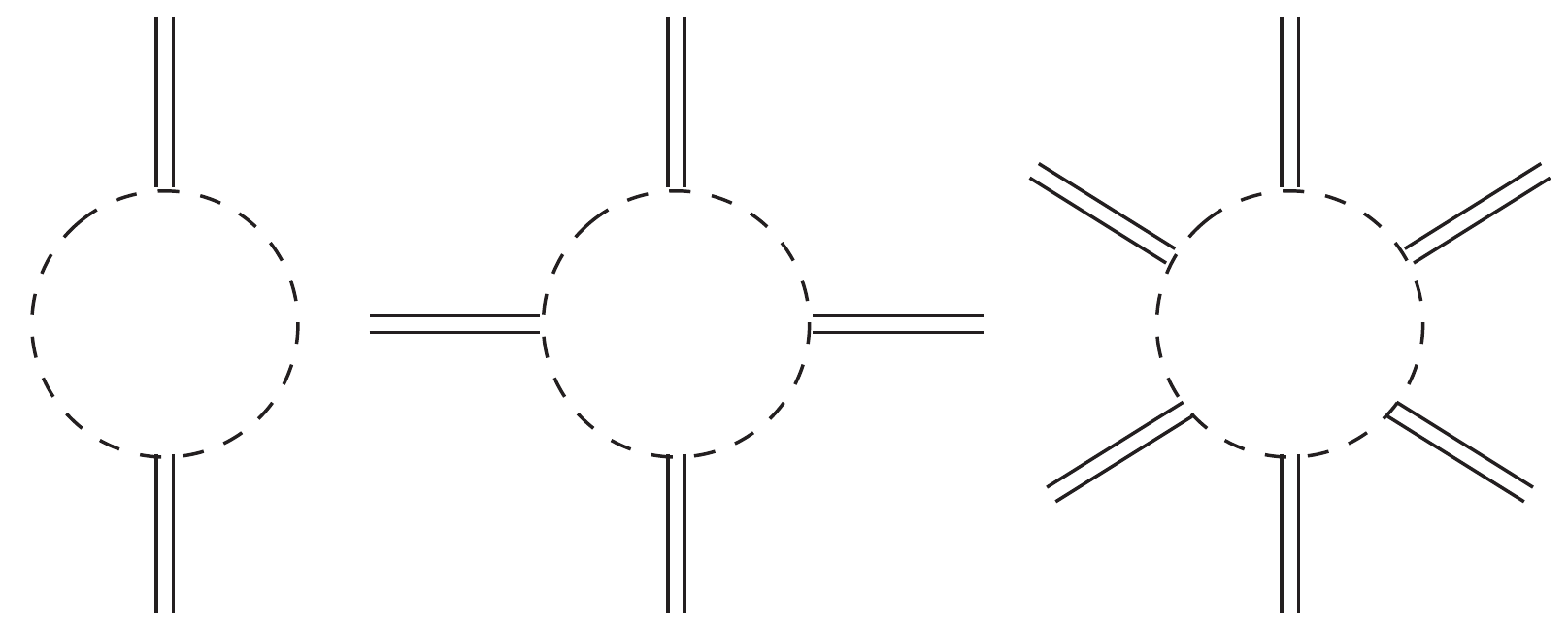}
\end{center}
\caption{One-loop supergraphs composed by propagators $\langle\psi(1)\bar\psi(2)\rangle$.}
\end{figure}

\begin{figure}[!h]
\begin{center}
\includegraphics[angle=0,scale=0.70]{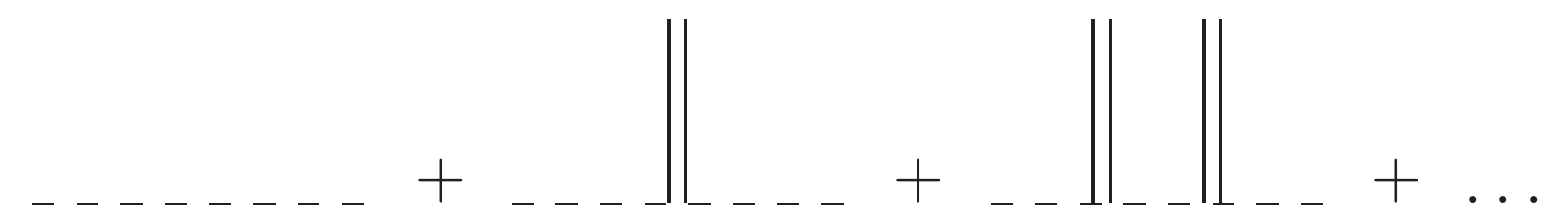}
\end{center}
\caption{Dressed propagator $\langle \psi(1)\bar\psi(2)\rangle_D$. The vertices are $\big[\frac{1}{4\beta}+(2h)^2(\Phi\bar{\Phi})\big]\psi\bar{\psi}$.}
\end{figure}

Now, let us sum over the vertices $\big[-\frac{1}{4}\big(1+\frac{1}{\beta}\big)+M_\psi\big]\psi^2$ and $\big[-\frac{1}{4}\big(1+\frac{1}{\beta}\big)+M_\psi\big]\bar\psi^2$. The corresponding supergraphs exhibit their structures in Fig. 6 with only even number of vertices. Since we can insert an arbitrary number of vertices $\big[\frac{1}{4\beta}+(2h)^2(\Phi\bar{\Phi})\big]\psi\bar\psi$ into the propagators $\langle\psi(1)\bar\psi(2)\rangle$, we must introduce the dressed propagator $\langle \psi(1)\bar\psi(2)\rangle_D$ (see Fig. 7). Therefore, this dressed propagator is equal to
\bea
\label{dressedpropsum}
\langle \psi(1)\bar\psi(2)\rangle_D&=&\langle \psi(1)\bar\psi(2)\rangle+\int d^4\theta_3\langle \psi(1)\bar\psi(3)\rangle[(2h)^2\bar\Phi\Phi]_3\langle \psi(3)\bar\psi(2)\rangle+\int d^4\theta_3d^4\theta_4\nonumber\\
&\times&\langle \psi(1)\bar\psi(3)\rangle[(2h)^2\bar\Phi\Phi]_3\langle \psi(3)\bar\psi(4)\rangle[(2h)^2\bar\Phi\Phi]_4\langle \psi(4)\bar\psi(2)\rangle+\ldots \ .
\eea
By using (\ref{propagator2}) and proceeding as above, we arrive at
\bea
\label{dressedpropagator1}
\langle \psi(1)\bar\psi(2)\rangle_D=\bigg(\frac{4\beta\Pi_{1/2}}{\beta-1-4\beta(2h)^2\bar\Phi\Phi}\bigg)_1\delta_{12} \ .
\eea
Afterwards, we can compute all the contributions by noting that each one-loop supergraph above is formed by $n$ vertices like those ones given by Fig. 8.

\begin{figure}[!h]
\begin{center}
\includegraphics[angle=0,scale=0.625]{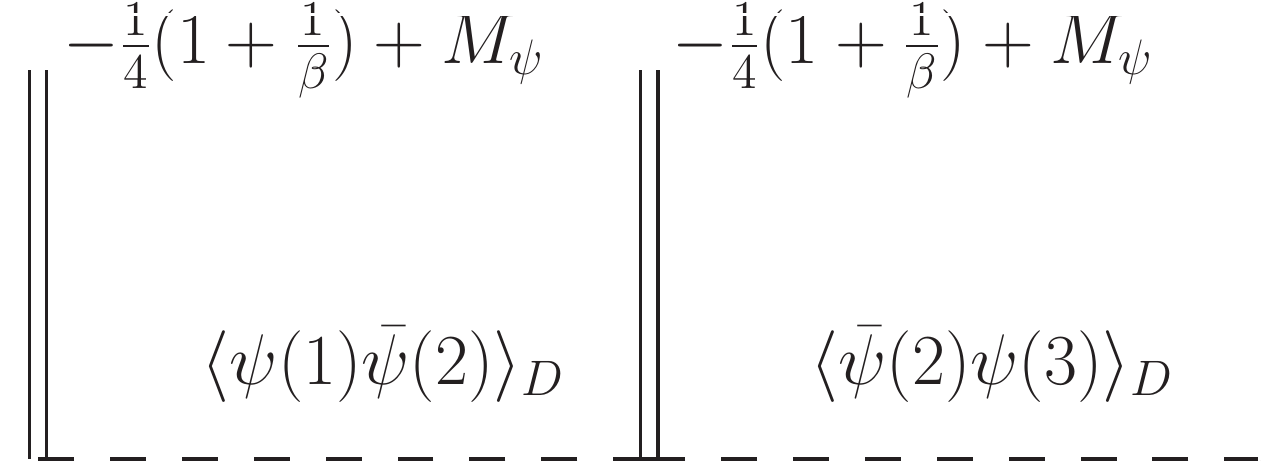}
\end{center}
\caption{A typical vertex in one-loop supergraphs involving $\big[-\frac{1}{4}\big(1+\frac{1}{\beta}\big)+M_\psi\big]\psi^2$ and $\big[-\frac{1}{4}\big(1+\frac{1}{\beta}\big)+M_\psi\big]\bar\psi^2$.}
\end{figure}

Hence, the contribution of this vertex is given by
\bea
L_{13}&=&\int d^4\theta_2\big[-\frac{1}{4}\big(1+\frac{1}{\beta}\big)+M_\psi\big]_1\Big[\Big(\frac{4\beta\Pi_{1/2}}{\beta-1-4\beta(2h)^2\bar\Phi\Phi}\Big)_1\delta_{12}\Big]\big[-\frac{1}{4}\big(1+\frac{1}{\beta}\big)+M_\psi\big]_2\nonumber\\
&\times&\Big[\Big(\frac{4\beta\Pi_{1/2}}{\beta-1-4\beta(2h)^2\bar\Phi\Phi}\Big)_2 \delta_{23}\Big]\nonumber\\
&=&\bigg(\frac{\beta+1-4\beta M_\psi}{\beta-1-4\beta(2h)^2\bar\Phi\Phi}\Pi_{1/2}\bigg)^2_1\delta_{13} \ .
\eea
It follows from the result above that the contribution of a supergraph formed by $n$ vertices is given by
\bea
(I_c)_n&=&\int d^4x\frac{1}{2n}\int d^4\theta_1d^4\theta_3\ldots d^4\theta_{2n-1}\int \frac{d^4p}{(2\pi)^4}L_{13}L_{35}\ldots L_{2n-3,2n-1} L_{2n-1,1} \nonumber\\
&=&\int d^8z\frac{1}{2n}\int \frac{d^4p}{(2\pi)^4}\bigg(\frac{\beta+1-4\beta M_\psi}{\beta-1-4\beta(2h)^2\bar\Phi\Phi}\Pi_{1/2}\bigg)^{2n}\Pi_{1/2}\delta_{\theta\theta^{\prime}}|_{\theta=\theta^{\prime}}\ .
\eea
On one hand, for $\beta=0$, we get ($\Pi_{1/2}\delta_{\theta\theta^{\prime}}|_{\theta=\theta^{\prime}}=2/p^2$)
\bea
(I_c)_n=\int d^8z\frac{1}{n}\int \frac{d^4p}{(2\pi)^4}\frac{1}{p^2}\ .
\eea
This integral over the momenta vanishes within the dimensional regularization scheme. Therefore, we get
\bea
\label{part3.1}
\Gamma^{(1)}_c=0 \ , \ \textrm{for $\beta=0$} \ .
\eea
On the other hand, for $\beta\ne0$, we obtain the effective action
\bea
\label{part3.2}
\Gamma^{(1)}_c=\sum_{n=1}^{\infty}(I_c)_n=-\int d^8z\int \frac{d^4p}{(2\pi)^4}\frac{1}{p^2}\ln\bigg\{1-\bigg[\frac{\beta+1-4\beta(-\frac{m_\psi^2}{2p^2}+(2h)^2\bar\Phi\Phi)}{\beta-1-4\beta(2h)^2\bar\Phi\Phi}\bigg]^{2}\bigg\} \ ,
\eea
for $\beta\ne0$. Moreover, we used $M_\psi\equiv-\frac{m_\psi^2}{2p^2}+(2h)^2\bar\Phi\Phi$. In particular, if $m_\psi=0$, Eq. (\ref{part3.2}) also vanishes within the dimensional regularization scheme.
\begin{figure}[!h]
\begin{center}
\includegraphics[angle=0,scale=0.475]{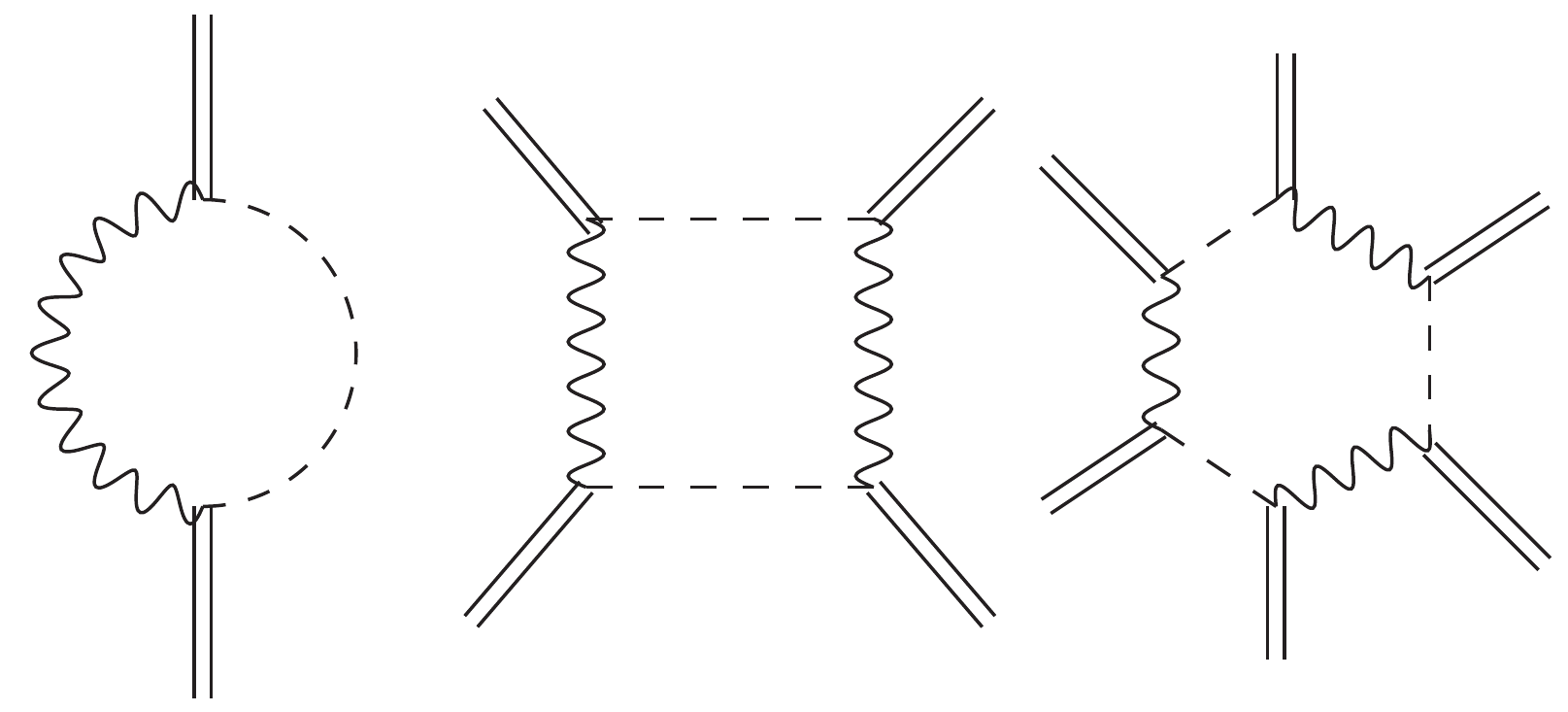}
\end{center}
\caption{One-loop supergraphs composed by propagators $\langle\psi(1)\bar\psi(2)\rangle$ and $\langle V(1)V(2)\rangle$.}
\end{figure}
\begin{figure}[!h]
\begin{center}
\includegraphics[angle=0,scale=0.725]{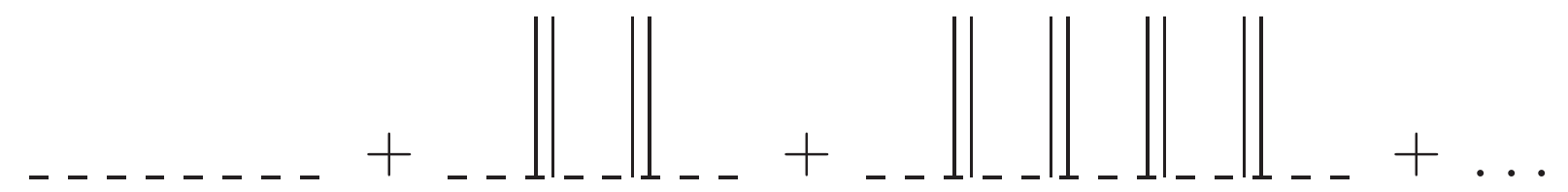}
\end{center}
\caption{Dressed propagator $\langle \psi(1)\bar\psi(2)\rangle_{2D}$.}
\end{figure}

Finally, let us evaluate the last type of one-loop supergraphs, which involve the propagators $\langle\psi(1)\bar\psi(2)\rangle$ and $\langle V(1)V(2)\rangle$ in the internal lines connecting the vertices $MV\psi$ and $MV\bar\psi$ (see Fig. 9). As before, we can insert an arbitrary number of vertices $\big[\frac{1}{4\beta}+(2h)^2(\Phi\bar{\Phi})\big]\psi\bar\psi$ into the propagators $\langle\psi(1)\bar\psi(2)\rangle$. Moreover, we can also insert an arbitrary number of pairs of the vertices $\big[-\frac{1}{4}\big(1+\frac{1}{\beta}\big)+M_\psi\big]\psi^2$ and $\big[-\frac{1}{4}\big(1+\frac{1}{\beta}\big)+M_\psi\big]\bar\psi^2$ into $\langle\psi(1)\bar\psi(2)\rangle$. Since $\langle\psi(1)\bar\psi(2)\rangle$ has already been dressed by $\big[\frac{1}{4\beta}+(2h)^2(\Phi\bar{\Phi})\big]\psi\bar\psi$ in (\ref{dressedpropsum}-\ref{dressedpropagator1}), it follows that the desired dressed propagator $\langle \psi(1)\bar\psi(2)\rangle_{2D}$ can be obtained through the summation over all pairs of the vertices $\big[-\frac{1}{4}\big(1+\frac{1}{\beta}\big)+M_\psi\big]\psi^2$ and $\big[-\frac{1}{4}\big(1+\frac{1}{\beta}\big)+M_\psi\big]\bar\psi^2$ into $\langle\psi(1)\bar\psi(2)\rangle_D$ (see Fig. 10). Therefore, we get
\bea
\langle \psi(1)\bar\psi(2)\rangle_{2D}&=&\langle \psi(1)\bar\psi(2)\rangle_D+\int d^4\theta_3d^4\theta_4\langle \psi(1)\bar\psi(3)\rangle_D\big[-\frac{1}{4}\big(1+\frac{1}{\beta}\big)+M_\psi\big]_3\langle \bar\psi(3)\psi(4)\rangle_D\nonumber\\
&\times&\big[-\frac{1}{4}\big(1+\frac{1}{\beta}\big)+M_\psi\big]_4\langle \psi(4)\bar\psi(2)\rangle_D+\int d^4\theta_3d^4\theta_4d^4\theta_5d^4\theta_6\langle \psi(1)\bar\psi(3)\rangle_D\nonumber\\
&\times&\big[-\frac{1}{4}\big(1+\frac{1}{\beta}\big)+M_\psi\big]_3\langle \bar\psi(3)\psi(4)\rangle_D\big[-\frac{1}{4}\big(1+\frac{1}{\beta}\big)+M_\psi\big]_4\langle \psi(4)\bar\psi(5)\rangle_D\nonumber\\
&\times&\big[-\frac{1}{4}\big(1+\frac{1}{\beta}\big)+M_\psi\big]_5\langle \bar\psi(5)\psi(6)\rangle_D\big[-\frac{1}{4}\big(1+\frac{1}{\beta}\big)+M_\psi\big]_6\langle \psi(6)\bar\psi(2)\rangle_D\nonumber\\
&+&\ldots \ .
\eea
After some algebraic work, we find
\bea
\langle \psi(1)\bar\psi(2)\rangle_{2D}=(f(\bar\Phi\Phi)\Pi_{1/2})_1\delta_{12} \ ,
\eea
where
\bea
\label{f}
f(\bar\Phi\Phi)\equiv\frac{[1-\beta(1-4(2h)^2\bar\Phi\Phi)]p^4}{(1-4(2h)^2\bar\Phi\Phi)p^4+[1+\beta(1-4(2h)^2\bar\Phi\Phi)]m_\psi^2p^2+\beta m^4_\psi} \ .
\eea
As before, we can compute all the contributions by noting that each supergraph above (Fig. 9) is formed by $n$ links depicted in Fig. 11, each of which yields the contribution
\bea
N_{13}&=&\int d^4\theta_2(M)_1\Big\{-\bigg[\frac{\Pi_{1/2}}{p^2+m^2_V+(2g)^2\bar\Phi\Phi}+\frac{\alpha\Pi_0}{p^2+\alpha\big(m^2_V+(2g)^2\bar\Phi\Phi\big)}\bigg]_1\delta_{12}\Big\}(M)_2\nonumber\\
&\times&\Big[(f\Pi_{1/2})_2\delta_{23}\Big]\nonumber\\
&=&\bigg(\frac{-fM^2\Pi_{1/2}}{p^2+m^2_V+(2g)^2\bar\Phi\Phi}\bigg)_1\delta_{13} \ .
\eea

\begin{figure}[!h]
\begin{center}
\includegraphics[angle=0,scale=0.625]{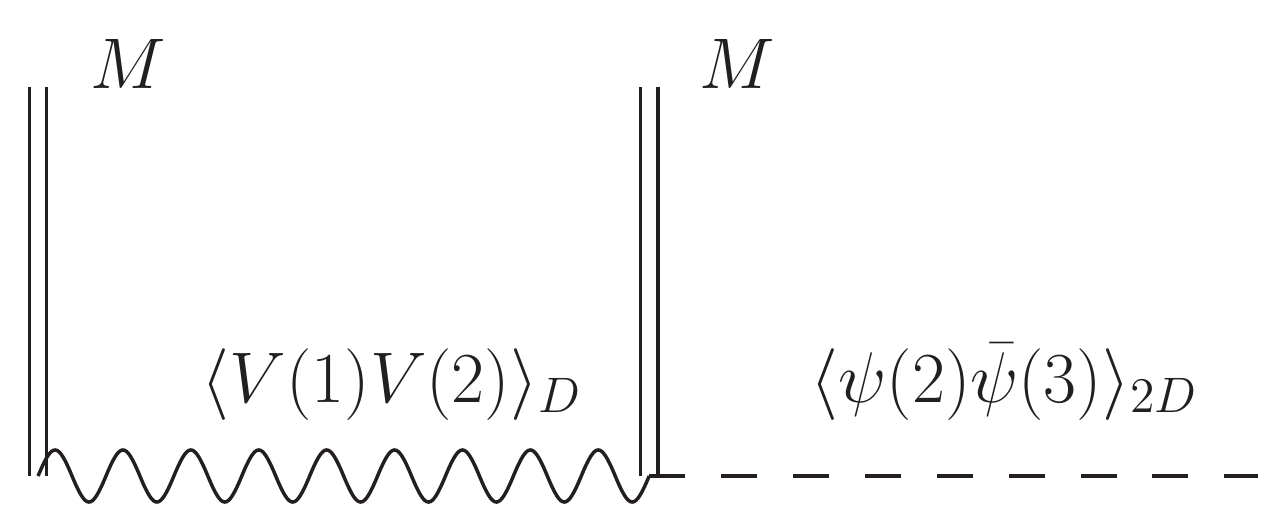}
\end{center}
\caption{A typical vertex in one-loop supergraphs involving $MV\psi$ and $MV\bar\psi$.}
\end{figure}

Hence, the contribution of a supergraph formed by $n$ such links is given by
\bea
(I_d)_n&=&\int d^4x\frac{1}{2n}\int d^4\theta_1d^4\theta_3\ldots d^4\theta_{2n-1}\int \frac{d^4p}{(2\pi)^4}N_{13}N_{35}\ldots N_{2n-3,2n-1} N_{2n-1,1} \nonumber\\
&=&\int d^8z\frac{1}{2n}\int \frac{d^4p}{(2\pi)^4}\bigg(\frac{-fM^2}{p^2+m^2_V+(2g)^2\bar\Phi\Phi}\bigg)^n\Pi_{1/2}\delta_{\theta\theta^{\prime}}|_{\theta=\theta^{\prime}}\ .
\eea
Again, by using $\Pi_{1/2}\delta_{\theta\theta^{\prime}}|_{\theta=\theta^{\prime}}=2/p^2$, we get the effective action
\bea
\label{part4}
\Gamma^{(1)}_d=\sum_{n=1}^{\infty}(I_d)_n=-\int d^8z\int \frac{d^4p}{(2\pi)^4}\frac{1}{p^2}\ln\bigg[1+\frac{fM^2}{p^2+m^2_V+(2g)^2\bar\Phi\Phi}\bigg] \ .
\eea

For $\beta=0$, we get the total one-loop KEP (up to terms independent on the background superfields) by summing (\ref{part1+2}), (\ref{part3.1}), and (\ref{part4})
\bea
\label{totalKEP1}
K^{(1)}_{\beta=0}(\bar\Phi\Phi)=-\int \frac{d^4p}{(2\pi)^4}\frac{1}{p^2}\ln\bigg\{p^2+m^2_V+(2g)^2\bar\Phi\Phi+\frac{M^2p^2}{(1-4(2h)^2\bar\Phi\Phi)p^2+m^2_\psi}\bigg\} ,
\eea
where we substituted the explicit form of $f$ for $\beta=0$.

For $\beta\ne0$, we obtain the total one-loop KEP by summing (\ref{part1+2}), (\ref{part3.2}), and (\ref{part4})
\bea
\label{totalKEP2}
&&K^{(1)}_{\beta\ne0}(\bar\Phi\Phi)=-\int \frac{d^4p}{(2\pi)^4}\frac{1}{p^2}\Bigg\{\ln\bigg\{1-\bigg[\frac{\beta+1-4\beta(-\frac{m_\psi^2}{2p^2}+(2h)^2\bar\Phi\Phi)}{\beta-1-4\beta(2h)^2\bar\Phi\Phi}\bigg]^{2}\bigg\}+\ln\bigg\{p^2\nonumber\\
&&+m^2_V+(2g)^2\bar\Phi\Phi+\frac{[1-\beta(1-4(2h)^2\bar\Phi\Phi)]M^2p^4}{(1-4(2h)^2\bar\Phi\Phi)p^4+[1+\beta(1-4(2h)^2\bar\Phi\Phi)]m^2_\psi p^2+\beta m^4_\psi}\bigg\}\Bigg\} \ ,
\eea
where we substituted the explicit form of $f$ (\ref{f}). Notice that (\ref{totalKEP2}) depends on the gauge parameter $\beta$, but it does not depend on the gauge parameter $\alpha$ ( one should remind that $\beta$ corresponds to the gauge fixing for the $\psi_{\alpha}$ field, and $\alpha$ -- for the real $V$ gauge field, and the gauge independence, that is, $\alpha$-independence of the one-loop KEP in the super-QED involving only chiral matter and $V$ field is a well-known fact \cite{SYM}).

Unfortunately we did not succeed to perform the momentum integrals (\ref{totalKEP2}) analytically and find an explicit expression for the $\beta$-dependent term in a most generic case. Therefore, in order to proceed with the calculation and solve explicitly the integral above at least in certain cases, we will consider two characteristic examples where the final result is expressed in closed form and in terms of elementary functions.

As our first example, let us take $m_\psi=0$ in (\ref{totalKEP1},\ref{totalKEP2}). It follows that
\bea
\label{particularKEP1}
&&K^{(1)}_{m_\psi=0}(\bar\Phi\Phi)=-\int \frac{d^4p}{(2\pi)^4}\frac{1}{p^2}\Bigg\{\ln\bigg\{1-\bigg[\frac{\beta+1-4\beta(2h)^2\bar\Phi\Phi}{\beta-1-4\beta(2h)^2\bar\Phi\Phi}\bigg]^{2}\bigg\}+\ln\bigg\{p^2\nonumber\\
&&+m^2_V+(2g)^2\bar\Phi\Phi+\frac{[1-\beta(1-4(2h)^2\bar\Phi\Phi)]M^2}{1-4(2h)^2\bar\Phi\Phi}\bigg\}\Bigg\} \ ,
\eea
The first integral in this expression vanishes within the dimensional reduction scheme. The second one is well known and, in the limit $\omega\rightarrow2$, gives
\bea
\label{mainresult1}
K^{(1)}_{m_\psi=0}=K_{m_\psi=0,div}^{(1)}(\bar\Phi\Phi)+K_{m_\psi=0,fin}^{(1)}(\bar\Phi\Phi) \ ,
\eea
where
\bea
\label{divresult1}
K_{m_\psi=0,div}^{(1)}(\bar\Phi\Phi)&=&-\frac{1}{16\pi^2(2-\omega)}\Big[m^2_V+(2g)^2\bar\Phi\Phi+\frac{[1-\beta(1-4(2h)^2\bar\Phi\Phi)]M^2}{1-4(2h)^2\bar\Phi\Phi}\Big],\\
\label{finresult1}
K_{m_\psi=0,fin}^{(1)}(\bar\Phi\Phi)&=&\frac{1}{16\pi^2}\Big[m^2_V+(2g)^2\bar\Phi\Phi+\frac{[1-\beta(1-4(2h)^2\bar\Phi\Phi)]M^2}{1-4(2h)^2\bar\Phi\Phi}\Big]\nonumber\\
&\times&\ln\frac{1}{\mu^2}\Big[m^2_V+(2g)^2\bar\Phi\Phi+\frac{[1-\beta(1-4(2h)^2\bar\Phi\Phi)]M^2}{1-4(2h)^2\bar\Phi\Phi}\Big] \ ,
\eea
$M=\frac{1}{2}(m-8gh\bar\Phi\Phi)$ and $\mu$ is an arbitrary scale required on dimensional grounds. If we take the particular case of $\beta=-1$ and $m_V=0$, we recover the result of Ref. \cite{prev}.

As our second example, let us consider $\beta=0$. Hence, we only need to calculate (\ref{totalKEP1}). The procedure to calculate it is quite analogous to the one reported in Ref. \cite{prev2}. Therefore, we get
\bea
\label{mainresult2}
K^{(1)}_{\beta=0}=K_{\beta=0,div}^{(1)}(\bar\Phi\Phi)+K_{\beta=0,fin}^{(1)}(\bar\Phi\Phi) \ ,
\eea
where
\bea
\label{divresult2}
K_{\beta=0,div}^{(1)}(\bar\Phi\Phi)&=&-\frac{1}{16\pi^2(2-\omega)}\Big[m^2_V+(2g)^2\bar\Phi\Phi+\frac{M^2}{1-4(2h)^2\bar\Phi\Phi}\Big],\\
\label{finresult2}
K_{\beta=0,fin}^{(1)}(\bar\Phi\Phi)&=&\frac{1}{16\pi^2}\Big[\Omega_+\ln\Big(\frac{\Omega_+}{\mu^2}\Big)+\Omega_-\ln\Big(\frac{\Omega_-}{\mu^2}\Big)
-\Omega_3\ln\Big(\frac{\Omega_3}{\mu^2}\Big)\Big] \ .
\eea
Moreover, we introduced a shorthand notation:
\bea
\label{omega}
\Omega_{\pm}&=&\frac{1}{2}\Big\{m^2_V+(2g)^2\bar\Phi\Phi+\frac{m_\psi^2+M^2}{1-4(2h)^2\bar\Phi\Phi}\nonumber\\
&\pm&\sqrt{\Big[m^2_V+(2g)^2\bar\Phi\Phi+\frac{m_\psi^2+M^2}{1-4(2h)^2\bar\Phi\Phi}\Big]^2
-4m_\psi^2(m^2_V+(2g)^2\bar\Phi\Phi)}\Big\}\nonumber\\
\Omega_{3}&=&\frac{m_\psi^2}{1-4(2h)^2\bar\Phi\Phi} \ .
\eea

Notice that one-loop results (\ref{mainresult1},\ref{mainresult2}) are both divergent. Moreover, we notice that the divergent parts (\ref{divresult1},\ref{divresult2}) are non-polynomial, and to eliminate the divergences, it would be necessary to introduce an infinite number of counterterms and an infinite number of unknown parameters in order to cancel the ultraviolet divergences appearing in the quantum corrections, so that the theory would not have any predictive power. However, it reflects the fact we already mentioned above, that theory under consideration is non-renormalizable and must be interpreted as an effective field theory for the low-energy domain \cite{Bur}. It is also clear, that in the case $h=0$, we notice that the divergent terms (\ref{divresult1},\ref{divresult2}) are proportional to $\bar\Phi\Phi$. Therefore, we can implement one-loop counterterm as the one used in the SQED \cite{SYM} to eliminate the divergences. However, in this case the coupling between chiral matter and chiral spinor gauge fields is switched off, therefore the spinor gauge field completely decouples, and the theory reduces to the usual super-QED.

\section{Summary}

We considered the Abelian superfield gauge theory involving two gauge fields, the real scalar one and the spinor one. The essentially new feature of this theory consists in the fact that it essentially involves the BF term thus opening the way for constructing of more sophisticated supersymmetric models involving the antisymmetric tensor fields.

The theory we consider represents itself as an alternative model involving two gauge fields, different from that one considered earlier in \cite{prev}. For this theory we calculated the one-loop K\"{a}hlerian effective potential which turns out to be divergent since the only possible gauge invariant coupling of the matter to the spinor gauge field turns out to be non-renormalizable. However, treating this theory as an effective model for description of low-energy limit of string theory (one can remind that the antisymmetric tensor field naturally emerges within the string context playing there an important role, see f.e. \cite{SW}), we can implement a natural cutoff of the order of the characteristic string mass.

Further application of this study could consist in development of supersymmetric extensions of more sophisticated theories involving the BF theory as an ingredient.

\vspace*{4mm}

{\bf Acknowledgments.} This work was partially supported by Conselho
Nacional de Desenvolvimento Cient\'{\i}fico e Tecnol\'{o}gico (CNPq). The work by A. Yu. P. has been supported by the
CNPq project No. 303438/2012-6. A. Yu. P. thanks the hospitality of the Universidade Federal do Cear\'{a} (UFC).

\end{document}